\documentclass[aps,prd,twocolumn,showpacs,superscriptaddress]{revtex4}
\newcommand{\PRE}[1]{}  

\usepackage{epsfig}
\usepackage{bm}
\newcommand{\be}{\begin{equation}}
\newcommand{\ee}{\end{equation}}
\newcommand{\beas}{\begin{eqnarray*}}
\newcommand{\eeas}{\end{eqnarray*}}
\newcommand{\bea}{\begin{eqnarray}}
\newcommand{\eea}{\end{eqnarray}}

\newcommand{\bwide}{\begin{widetext}}
\newcommand{\ewide}{\end{widetext}}
\newcommand{\postscript}[2]{\setlength{\epsfxsize}{#2\hsize}
   \centerline{\epsfbox{#1}}}

\newcommand{\gev}{\text{GeV}}
\newcommand{\tev}{\text{TeV}}

\newcommand{\cm}{\text{cm}}
\newcommand{\km}{\text{km}}

\newcommand{\s}{\text{s}}
\newcommand{\yr}{\text{yr}}
\newcommand{\sr}{\text{sr}}

\newcommand{\eqref}[1]{Eq.~(\ref{#1})}
\newcommand{\eqsref}[2]{Eqs.~(\ref{#1}) and (\ref{#2})}

\newcommand{\figref}[1]{Fig.~\ref{fig:#1}}

\newcommand{\phinu}{\phi^{\nu}}
\newcommand{\phinumax}{\phi^{\nu}_{\text{max}}}
\newcommand{\phiWB}{\phi^{\nu}_{\text{WB}}}
\newcommand{\phiAGN}{\phi^{\nu}_{\text{AGN}}}
\newcommand{\sigmanuN}{\sigma_{\nu N}}
\newcommand{\sigmaSM}{\sigma_{\text{SM}}}
\newcommand{\Ndown}{{\cal N}_{\text{down}}}
\newcommand{\Nup}{{\cal N}_{\text{up}}}
\newcommand{\Ndownobs}{\Ndown^{\text{obs}}}
\newcommand{\Nupobs}{\Nup^{\text{obs}}}
\newcommand{\nT}{N_{\text{T}}}

\begin{document}

\preprint{
\hfil
\begin{minipage}[t]{3in}
\begin{flushright}
\vspace*{.4in}
NUB--3255--Th--05 \\
UCI--TR--2005--13 \\
hep-ph/0504228
\end{flushright}
\end{minipage}
}

\title{
\PRE{\vspace*{1.5in}}
Particle Physics on Ice: \\
Constraints on Neutrino Interactions Far Above the Weak Scale
\PRE{\vspace*{0.3in}}
}

\author{Luis A.~Anchordoqui}
\affiliation{Department of Physics,
Northeastern University, Boston, MA 02115
\PRE{\vspace*{.1in}}
}

\author{Jonathan L.~Feng}
\affiliation{Department of Physics and Astronomy,
University of California, Irvine, CA 92697
\PRE{\vspace*{.1in}}
}

\author{Haim Goldberg}
\affiliation{Department of Physics,
Northeastern University, Boston, MA 02115
\PRE{\vspace*{.1in}}
}


\begin{abstract}
\PRE{\vspace*{.1in}} Ultra-high energy cosmic rays and neutrinos probe
energies far above the weak scale.  Their usefulness might appear to
be limited by astrophysical uncertainties; however, by simultaneously
considering up- and down-going events, one may disentangle particle
physics from astrophysics. We show that present data from the AMANDA
experiment in the South Pole ice already imply an upper
bound on neutrino cross sections at energy scales that will likely
never be probed at man-made accelerators. The existing data also place
an upper limit on the neutrino flux valid for any neutrino cross section.
In the future, similar analyses of IceCube data will
constrain neutrino properties and fluxes at the ${\cal O}(10\%)$
level.
\end{abstract}

\pacs{96.40.Tv, 04.50.+h \hspace{4.2cm} NUB--3255--Th--05; UCI--TR--2005--13}

\maketitle

Ultra-high energy cosmic rays have been observed with energies above
$10^{10}~\gev$, implying collisions with terrestrial protons at
center-of-mass energies above 100 TeV.  Ultra-high energy cosmic
neutrinos, so far undetected, are expected to accompany these cosmic
rays.  These cosmic neutrinos are especially interesting, because
their known interactions are so weak that they are highly sensitive to
new interactions at energies far above the weak scale, where new
interactions are expected in many extensions of the standard model (SM) of
particle physics.  In addition, ultra-high energy neutrinos are unique
messengers, as they are expected to escape from (and point back to)
even the most dense astrophysical sources.

The promise of ultra-high energy neutrinos might appear to be severely
limited by astrophysical uncertainties.  Event rates constrain only a
combination of fluxes and cross sections, and so astrophysical
uncertainties cloud particle physics implications and vice versa.
However, the event rates for up- and down-going neutrinos depend
differently on neutrino cross
sections~\cite{Kusenko:2001gj,Feng:2001ib}.  By combining both up- and
down-going data one may therefore disentangle particle physics from
astrophysics and constrain both the properties of astrophysical
sources and the interactions of neutrinos far above the weak scale.

Here we consider neutrino telescopes operating under ice at the South
Pole~\cite{Andres:1999hm,Ahrens:2002dv}. We show that current data from the
AMANDA South Pole telescope~\cite{Ackermann:2005sb}
already significantly constrain neutrino cross sections at
center-of-mass energies $\sqrt{s}\approx 6~\tev$, and future data will
provide ${\cal O} (10\%)$ determinations.  These results will be
complemented~\cite{Anchordoqui:2001cg} by future data from the Pierre
Auger Observatory~\cite{Abraham:2004dt}.  The energies probed are far
above the HERA domain $\sqrt{s}\simeq 500~\gev$, the highest
accelerator energy at which even indirect tests of neutrino cross
sections have been made.

Simple parton model predictions for neutrino-nucleon cross
sections~\cite{Gandhi:1998ri} may be suppressed, for example, by
saturation effects at small $x$~\cite{Gribov:1984tu}.  Such effects
have been proposed to slow down the power law scaling of cross section
with neutrino energy to comply with the Froissart
bound~\cite{Kwiecinski:1990tb}.  On the other hand, neutrino cross
sections may also be enhanced, for example, by the exchange of towers
of Kaluza-Klein gravitons~\cite{Nussinov:1998jt}, black hole
production~\cite{Feng:2001ib,Banks:1999gd},  TeV-scale string
excitations~\cite{Domokos:1998ry}, or electroweak instanton
processes~\cite{Fodor:2003bn}. Our results constrain all of these
possibilities.

The results derived below also constrain the extraterrestrial neutrino
flux, which is at present unknown. Neutrino sources may be
conveniently characterized as either optically thin or thick.  In
optically thin sources, the nucleons responsible for neutrino
production (through photoproduction or $pp$ collisions) escape the
source and constitute the observed cosmic rays. The observed cosmic
ray and diffuse neutrino fluxes are then
correlated~\cite{Waxman:1998yy}.  In contrast, in optically thick
sources, the neutrino progenitors do not escape, thus vitiating the
relation between cosmic ray and neutrino
fluxes~\cite{Mannheim:1998wp}. In principle, this permits a very large
enhancement to the neutrino flux. We derive upper bounds on this
enhancement that are valid for any neutrino cross section.

We will derive bounds without assuming particular neutrino fluxes or
cross sections.  It will be convenient, however, to present results
relative to standard reference values.  For the reference cross
section, we choose $\sigmaSM$, the charged current (CC)
neutrino-nucleon cross section of the simple parton
model~\cite{Gandhi:1998ri}.  For the reference flux, we adopt the
Waxman-Bahcall (WB) flux $\phiWB$~\cite{Waxman:1998yy}.  This flux is
that of an optically thin source. At production, the WB flux has flavor
ratios $\nu_\mu:\nu_e:\nu_\tau = 2: 1: 0$, but this quickly transforms
to $1:1:1$ through neutrino oscillations, and so $E^2 \, \phiWB \simeq
2 \times 10^{-8}~\gev~\cm^{-2}~\s^{-1}~\sr^{-1}$ for each flavor.
In what follows, we focus on neutrino energies in the range
$10^7~\gev$ to $10^{7.5}~\gev$, where the background from atmospheric
neutrinos is negligible, but the extraterrestrial flux is expected to
be significant. Hereafter, we take $\langle E \rangle = 10^{7.25}$~GeV.

Neutrinos are detected in neutrino telescopes when they create charged
leptons or showers through CC or neutral current (NC) events near the
neutrino telescope, and the resulting leptons or showers propagate
into the experiment's instrumented volume.  For down-going events, the
probability of neutrino conversion is always small, barring
extraordinary enhancements to neutrino cross sections.  Letting
$\sigmanuN$ denote the total (NC $+$ CC) neutrino-nucleon cross
section, the down-going event rate is therefore
\begin{equation}
\Ndown = C_{\text{down}} \ \frac{\phinu}{\phiWB}
\ \frac{\sigmanuN}{\sigmaSM} \ ,
\label{down}
\end{equation}
where $\phinu$ indicates the average extraterrestrial
$\nu + \overline \nu$ flux per flavor in the energy bin of interest.
The constant $C_{\text{down}}$ depends on exposure and acceptance and
varies according to neutrino flavor from experiment to experiment.

For up-going events, the dependence on cross section is completely
different.  At energies above $10^6~\gev$, neutrino interaction
lengths become smaller than the radius of the Earth.  For $\nu_e$
and $\nu_\mu$, this implies that most upward-going neutrinos are
shadowed by the Earth, and only those that are
Earth-skimming~\cite{Feng:2001ue}, traveling at low angles along
chords with lengths of order their interaction length, can produce
a visible signal.  For $\nu_\tau$, this attenuation is somewhat
offset by effects of regeneration $\nu_\tau \to \tau \to
\nu_\tau$~\cite{Halzen:1998be}. However, in the energy range of
interest, the expected events will be largely dominated by
neutrinos surviving until the region of the detector.

The dependence of upward-going event rates on anomalous neutrino cross
sections depends on the source of the anomaly.  We consider two
prominent cases.  First, in many new physics cases, the SM
CC cross section $\sigmaSM$ is not altered, but there are new neutrino
interactions that produce showers.  Letting $\sigmanuN$ be the total
neutrino cross section, including the standard model CC and new
physics contributions, the up-going event rate
is~\cite{Anchordoqui:2001cg}
\begin{equation}
\Nup = C_{\text{up}} \ \frac{\phinu}{\phiWB}
\ \frac{\sigmaSM^2}{\sigmanuN^2}
\quad \left(\frac{\sigmanuN}{\sigmaSM} > 1 \right) \ .
\label{up}
\end{equation}
As discussed in Ref.~\cite{Anchordoqui:2001cg}, \eqref{up} is valid
assuming $L^{l} \ll L^{\nu} < R_{\oplus}$,
where $L^{l}$ is the typical lepton path length in Earth, $L^{\nu}$ is the neutrino
interaction length in Earth, and $R_{\oplus}$ is the radius of the
Earth. This corresponds to  $E> 10^7~{\rm GeV}.$ Here, $\sigmanuN$ indicates
any enhancement of
the cross section which will increase the event
rate for down-going neutrinos, but because of absorption will
suppress the up-coming events. The latter can be achieved through
cuts on shower energy fraction greater than or equal to that
characterizing the CC SM process. Extreme enhancements
to $\sigmanuN$ may reduce $L^{\nu}$ to
$L^l$, leading to a different parametric dependence in
\eqref{up}, but we neglect such cases here.  Second, we consider the
possibility of screening, in which both the standard model NC and CC
cross sections are reduced equally.  In this
case~\cite{Kusenko:2001gj,Anchordoqui:2001cg},
\begin{equation}
\Nup = C^{\text{screen}}_{\text{up}}
\ \frac{\phinu}{\phiWB} \ \frac{\sigmaSM}{\sigmanuN} \ ,
\label{upscreen}
\end{equation}
where $\sigmanuN$ and $\sigmaSM$ are CC cross sections with and
without screening, respectively.

Given the parametric dependences of Eqs.~(\ref{down}), (\ref{up}), and
(\ref{upscreen}), we now consider the implications of existing data
from AMANDA.  Assuming the standard model CC neutrino interaction, the
AMANDA Collaboration has derived the 90\% CL upper bound on the
diffuse neutrino flux $E^2\,\phinumax = 3.3 \times
10^{-7}~\gev~\cm^{-2}~\s^{-1}~\sr^{-1}$ per
flavor~\cite{Ackermann:2005sb}, assuming an $E^{-2}$ dependence of the
flux, valid for $10^6~\gev$ to $10^{9.5}~\gev$ neutrinos. Since the energy
distribution of the AMANDA data peaks in the energy bin of interest, it is
reasonable to use $\phinumax (\langle E \rangle)$ as the upper limit
in the bin. Here we generalize this bound to the case in which there are
{\em two} unknown
quantities: the neutrino cross section and the neutrino flux.  Applied
to $\nu_e$ down-going events, the constraint implies
\begin{equation}
\phinu  \ \frac{\sigmanuN}{\sigmaSM} < \phinumax \ . \label{Ad}
\end{equation}
Dividing \eqref{Ad} by $\phiWB$ gives
\begin{equation}
\frac{\phinu}{\phiWB} \ \frac{\sigmanuN}{\sigmaSM} < 16
\label{downconstraint}
\end{equation}
at 90\% CL. A similar analysis for up-going events yields
\begin{equation}
\frac{\phinu}{\phiWB} \ \frac{\sigmaSM^2}{\sigmanuN^2} < 16
\quad \left(\frac{\sigmanuN}{\sigmaSM} > 1 \right)
\label{upconstraint}
\end{equation}
for the case of new physics contributions, and
\begin{equation}
\frac{\phinu}{\phiWB} \ \frac{\sigmaSM}{\sigmanuN} < 16
\label{upconstraintscreen}
\end{equation}
for the case of screening.

\begin{figure}
\postscript{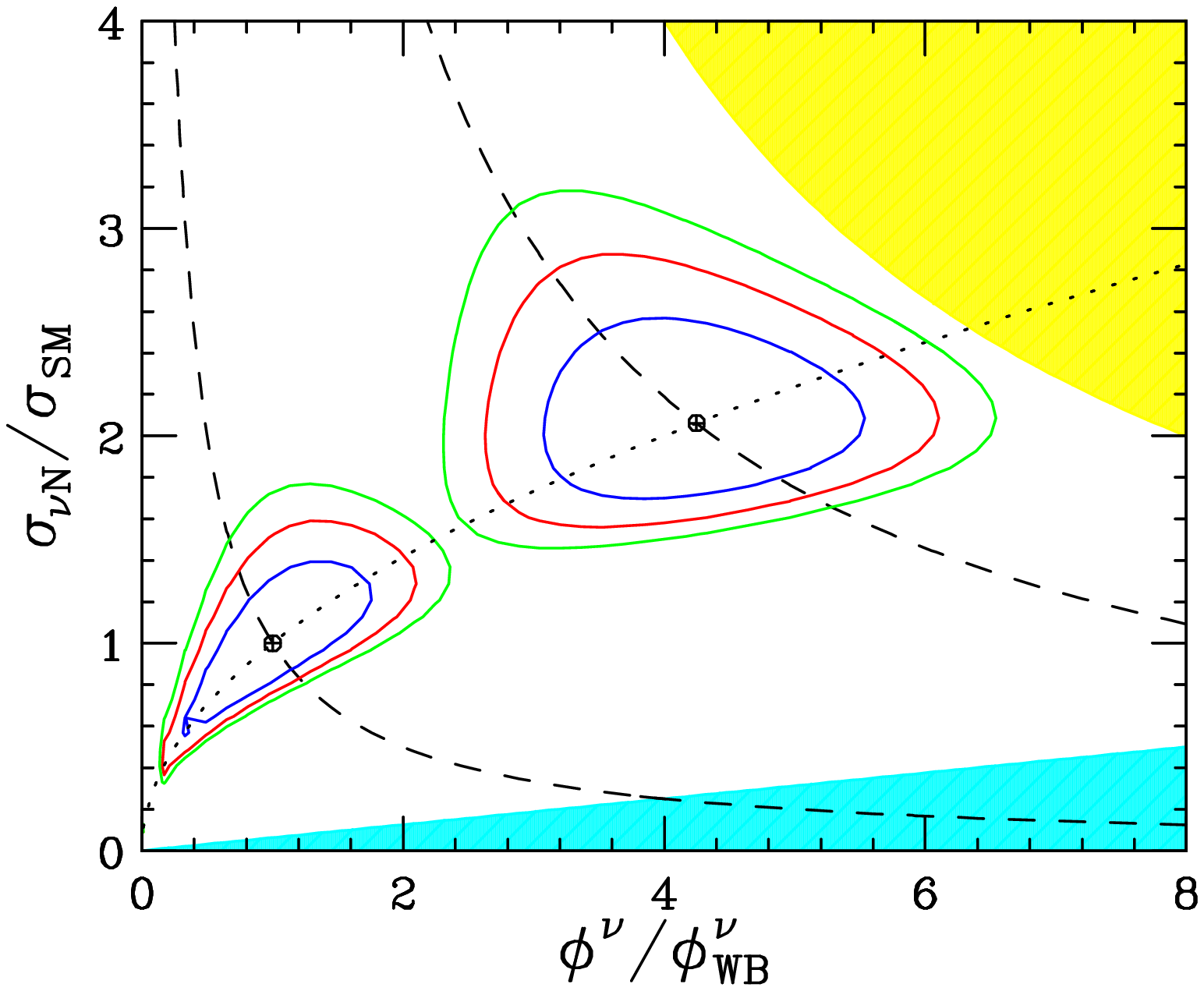}{0.98} \caption{Projected determinations
  of neutrino fluxes and cross sections at $\sqrt{s} \approx 6~\tev$
  from future IceCube data.  Two cases with $(\Ndownobs, \Nupobs) =
  (4,20)$ and $(35,20)$ are considered.  In both cases, the best fit
  flux and cross sections are shown, along with the 90\%, 99\%, and
  99.9\% CL inclusion contours.  Contours of constant $\Nup = 20$
  (dotted), $\Ndown = 4$ (left dashed), and $\Ndown = 35$ (right
  dashed) are also shown. Neutrino fluxes and cross sections excluded by AMANDA at 90\%CL
  are also indicated:  The upper (lower) shaded region is excluded by null results for
  down-going (up-going) events~\cite{Ackermann:2005sb}.}
\label{fig:IceCube}
\end{figure}

These constraints exclude the shaded regions of \figref{IceCube}.
The upper region is
excluded by down-going data and the lower region is excluded by the
up-going data, assuming screening.  The upper and lower regions meet
at $\sigmanuN / \sigmaSM =1$.  As a result, for any neutrino cross section,
we find an upper bound on the neutrino flux in
the energy range $10^7~\gev$ to $10^{7.5}~\gev$ of $\phinu < 16
\phiWB$.  Note that the lower shaded region limits the amount by which
screening effects can suppress $\sigmaSM$.

How will these results improve in the near future?  We now
consider the possible implications of IceCube, the successor
experiment to AMANDA. In the energy range of interest the $\tau$
decay length is comparable to the instrumented volume; thus, one
can observe all $\nu_\tau$ topologies: a $\tau$ track followed by
the $\tau$-decay shower (``lollipop''), a hadronic shower followed
by a $\tau$ track which leaves the detector (``popillol''), and
double bang events. All of these distinctive topologies allow a
direct and precise measurement of the incoming neutrino
energy~\cite{Halzen:2005qu}. We also note that the absence of
oscillation precludes a $\nu_\tau$ atmospheric background. For
these reasons, the detection prospects for up-going neutrinos are
brighter for $\nu_{\tau}$ than the other flavors, and we focus on
them below.

To evaluate the prospects for IceCube, we must determine
$C_{\text{down}}$ and $C_{\text{up}}$ for IceCube.  We focus on the
case in which new neutrino interactions modify the NC cross sections,
but leave the CC cross sections invariant.  In NC processes most of
the energy is carried off by pions. At TeV energies, the interaction
mean free path of $\pi^\pm$ in ice is orders of magnitude smaller than
the pion decay length, and so nearly all energy is channeled into
electromagnetic modes through $\pi^0$ decay. To estimate the
efficiencies for down-going events in the NC channel, we therefore
adopt as our basis of comparison the electromagnetic showers induced
by $\nu_e$~\cite{background}. With this in mind,
\begin{equation}
C_{\text{down}} = 2 \pi \phiWB \sigmaSM \nT T
   \Delta E \int_0^1 \eta(\cos\theta)\ d \cos\theta \ ,
\label{cdown}
\end{equation}
where $\nT$ is the number of target nucleons in the effective volume,
$0.5 <\eta(\cos\theta) < 1$ is the experimental efficiency for detection of
electron neutrino showers with zenith angle $\theta$, $T$ is the
running time, and $\Delta E = 2.2 \times 10^7~\gev$.

IceCube is now under construction,  in the process of growing to
its final size. Including absorption effects, it will have an 
effective aperture $(A\Omega)_{\text{eff}} \sim  \pi/2~\km^2~\sr$ for detecting
$\nu_{\tau}$~\cite{Ahrens:2002dv}.  The detector will consist of
80 kilometer-length strings, each instrumented with 60 optical
modules.  The number of target nucleons may therefore be estimated
to be $\nT \simeq 6 \times 10^{38}$. To remain conservative, we
take $\int_0^1 \eta(\cos \theta) \ d \cos\theta = 0.8$.  Inserting
these numbers into \eqref{cdown}, along with a lifetime of the
experiment $T = 15~\yr$ and $\sigmaSM ( \langle E \rangle ) \simeq
2 \times 10^{-33}$~cm$^{2}$~\cite{Gandhi:1998ri}, we obtain
$C_{\text{down}} \simeq 4$.  The corresponding quantity for
up-going events in our energy interval is $C_{\text{up}} = (A
\Omega)_{\text{eff}} \ T \int \phi^\tau(E) dE \simeq 20$, where
$\int \phi^\tau(E) dE = 8.5 \times
10^{-1}~\km^{-2}~\yr^{-1}~\sr^{-1}$ is the $\tau$-lepton flux
emerging from the Earth due to (unabsorbed) $\nu_\tau$ interactions inside the
Earth for incoming $\phinu = \phiWB$ and $\sigmanuN =
\sigmaSM$~\cite{Tseng:2003pn}.

Given these estimates of $C_{\text{down}}$ and $C_{\text{up}}$, we now
determine projected sensitivities of IceCube to neutrino fluxes and
cross sections.  For a given set of observed rates $\Nupobs$ and
$\Ndownobs$, two curves are obtained in the two-dimensional parameter
space by setting $\Nupobs = \Nup$ and $\Ndownobs = \Ndown$.  These
curves intersect at a point, yielding the most probable values of flux
and cross section for the given observations.  Fluctuations about this
point define contours of constant $\chi^2$ in an approximation to a
multi-Poisson likelihood analysis~\cite{Baker:1983tu}.

In \figref{IceCube}, we show results for two representative cases that
are consistent with the AMANDA bounds derived above.  In the first
case, we assume $\sigmanuN = \sigmaSM$ and $\phinu = \phiWB$, leading
to 4 down-going and 20 up-going events.  The 90\%, 99\%, and 99.9\% CL
contours are those given in the lower left of the figure.  (These
contours will be slightly distorted for $\sigmanuN/\sigmaSM < 1$,
where \eqref{up} receives corrections, but we neglect this effect.)
We see that, even in the case that event rates are in accord with
standard assumptions, the neutrino-nucleon cross section is bounded to
be within 40\% of the SM prediction at 90\% CL. This is at a
center-of-mass energy $\sqrt{s}\simeq 6$ TeV, far beyond the reach of
any future man-made accelerator~\cite{note1}.  In the second case, we consider a
scenario in which the number of observed upcoming events remains at
20, but the number of down-going events is 35. In the second case,
clearly one has discovered new physics at well beyond 5$\sigma$.

It is noteworthy that in the energy bin of interest, the
predicted~\cite{Kalashev:2002kx} diffuse flux of neutrinos from a
uniform distribution of blazars, $\phiAGN(E)$, is about 9 times
larger than the WB flux, i.e., $\phiAGN (\langle E \rangle) \simeq 9
\phiWB$. These neutrinos are expected to be produced in optically
thick cores of blazars when ultra-high energy protons scatter off the
accretion disk orbiting the active galactic nucleus
(AGN)~\cite{Neronov:2002se}. By re-scaling the integrated luminosity
in \eqsref{down}{up}, it is straightforward to see that {\em if the
extraterrestrial flux is at the $\phiAGN$ level, in 2 yr of operation
IceCube will probe 40\% (70\%) enhancements from SM predictions at the
90\% (99.9\%) CL}.

Finally, we note that the AMANDA constraints have significant
consequences for what may be seen at IceCube.  Substituting
\eqref{downconstraint} into \eqref{down} with $C_{\text{down}}$
determined for IceCube, the $\nu_e$ down-going event rate in the
energy range $10^7~\gev$ to $10^{7.5}~\gev$ is constrained to $\Ndown
< 4~\yr^{-1}$ at 90\% CL~\cite{note2}. Event rates at IceCube for low scale gravity
models have been presented~\cite{Alvarez-Muniz:2001mk}. These event
rates, along with this bound on $\Ndown$, can be used to constrain the
multidimensional Planck scale.

In summary, we have shown that the sensitivity of neutrino telescopes
in the Antarctic ice to both up- and down-going ultra-high energy
neutrinos provides a powerful probe of ultra-high energy neutrino
fluxes and anomalous neutrino interactions.  Current null results from
AMANDA already provide interesting constraints on the flux-cross
section parameter space.  First, these results constrain both
suppressions of the neutrino flux from putative screening effects and
enhancements from new physics. Second, they exclude large fluxes
$\phinu > 16 \phiWB$ for neutrino energies between $10^7~\gev$ and
$10^{7.5}~\gev$, for any neutrino cross section. These
energies correspond to neutrino-nucleon collisions at $\sqrt{s} \approx
6~\tev$, far above the weak scale and likely never to be accessible at
man-made accelerators.

In the future, IceCube will be able to determine both neutrino fluxes
and cross sections with impressive accuracy.  We have considered
neutrinos with energies between $10^7~\gev$ and $10^{7.5}~\gev$.  For
standard model cross sections and the WB flux, 40\% (70\%)
enhancements from standard model predictions may be excluded at 90\%
(99.9\%) CL in 15 years of running. Should the neutrino flux be at the
level predicted for optically thick blazars, these bounds can be
attained after two years of data collection at IceCube.  Our analysis
assumes an extraterrestrial neutrino flux with $1:1:1$ flavor
composition.  If IceCube finds a different flavor
mix~\cite{Learned:1994wg}, it will not be difficult to repeat this
analysis for the correct flavor ratios.

{\em Acknowledgments.} We thank Steve Barwick, Hallsie Reno and Doug
Cowen for discussion.  The work of LAA is supported in part by NSF
grant No.~PHY--0140407.  The work of JLF is supported in part by NSF
CAREER grant No.~PHY--0239817, NASA Grant No.~NNG05GG44G, and the
Alfred P.~Sloan Foundation.  The work of HG is supported in part by
NSF grant No.~PHY--0244507.


\begin{thebibliography}{99}
\bibitem{background} The $\nu_e$ atmospheric background is taken to be zero.
\bibitem{note2} To derive this upper limit we set $\cos (\theta) = 1.$
\end{thebibliography}
\end{document}